%% file: main.tex
\documentclass{article}

\PassOptionsToPackage{numbers,sort&compress}{natbib}
\usepackage[final]{neurips_2026}
\makeatletter
\renewcommand{\@noticestring}{Workshop on Who Verifies the Agents? Toward Reliable Agent Development, NeurIPS 2026.}
\makeatother

\usepackage[utf8]{inputenc}
\usepackage[T1]{fontenc}
\usepackage{amsmath,amssymb}
\usepackage{graphicx}
\usepackage{booktabs}
\usepackage{xcolor}
\usepackage{microtype}
\usepackage{enumitem}
\usepackage{tikz}
\usetikzlibrary{arrows.meta, positioning, fit, backgrounds, calc}
\definecolor{cprompt}{HTML}{2B6CB0}
\definecolor{ccode}{HTML}{2F855A}
\definecolor{ctools}{HTML}{DD6B20}
\definecolor{cfp}{HTML}{6B46C1}
\usepackage{hyperref}
\hypersetup{
  colorlinks=true,
  linkcolor=black,
  citecolor=blue!50!black,
  urlcolor=blue!50!black,
  pdftitle={Per-Component Skill Identity: A Rewrite-Stable Primitive for Agent Verification},
  pdfauthor={Hongliang Liu, Yuhao Wu, Tung-Ling Li},
}

\title{Per-Component Skill Identity:\\
A Rewrite-Stable Primitive for Agent Verification}

\author{%
  Hongliang Liu\thanks{Equal contribution.} \quad
  Yuhao Wu\footnotemark[1] \quad
  Tung-Ling Li\footnotemark[1] \\
  Palo Alto Networks \\
  \texttt{\{honliu, yuhwu, tuli\}@paloaltonetworks.com}
}

\begin{document}
\maketitle

\input{sections/abstract}
\input{sections/introduction}
\input{sections/method}
\input{sections/results}
\input{sections/robustness}
\input{sections/decomposition}
\input{sections/twoaxis}
\input{sections/related}
\input{sections/limitations}
\input{sections/conclusion}

\begin{ack}
We thank Hui Gao and Badar Ahmed for their support. This work was conducted at Palo Alto
Networks.
\end{ack}

\bibliographystyle{plainnat}
\bibliography{references}

\input{sections/appendix}

\end{document}

%% file: sections/abstract.tex
\begin{abstract}
AI agents acquire and execute \emph{skills} at runtime: bundles of prompt
instructions, executable code, and tool declarations fetched from marketplaces and,
increasingly, authored by other agents. Verifying such a population presupposes something
verification does not supply: a stable notion of skill identity for a verdict to
attach to, one that survives rewriting yet still separates distinct skills. Cryptographic
hashing is engineered to destroy exactly that similarity, since a one-character edit
scrambles the digest. We present a locality-sensitive fingerprint that embeds each component
of a skill and projects it to bits with a multi-bank SimHash, giving a signature of at most
120 bytes, independent of skill size, compared in constant time by Hamming distance. Keeping the fingerprint as
a \emph{per-component triple} (prompt, code, tools) instead of a single score makes it
readable: the triple shows which component carries the reuse or the change. On a constructed corpus it tracks the embedding it approximates to within $0.03$ AUC
at $77\times$ fewer bits. Across ten rewrite transforms up to cross-language translation, at
each method's own zero-observed-false-positive threshold it recovers all $50$ rewrites
($49$--$50$ at its a-priori threshold) where a lexical baseline recovers $45$, the difference
being five full rewrites of both components. We then measure
where the identity axis ends. On a $906$-skill injection benchmark the fingerprint recognizes
$93\%$ of generated injections as tampered copies of a registered base and localizes
the change mostly to the component each generator edited, yet only $3$ of $44$ in-the-wild malicious
skills resemble any benign skill. As a safety detector it reaches $F_1$ $0.735$ even at an
in-sample-optimal threshold, against behavioral verification's $0.946$. Recognition is not
trust: structural identity tells a verifier which verified artifact to compare against, while
behavior decides whether the result is safe.
\end{abstract}

%% file: sections/introduction.tex
\section{Introduction}
\label{sec:intro}

Large language model agents have moved from answering questions to acquiring and executing
\emph{skills}: reusable bundles of prompt instructions, executable code, and tool or API
declarations that an agent loads at runtime to extend what it can
do~\citep{yao2022react,schick2023toolformer}. Skills arrive from many places at once: public
marketplaces, internal hubs, partner registries, the Model Context Protocol
ecosystem~\citep{mcp}, and increasingly other agents that author skills automatically.
Skills also change in place. When an agent's prompt is rewritten, a tool is swapped, or its
code refactored, its author needs to know which of those moved before asking whether the
agent got better.

Verifying such a population runs into an identity problem before it runs into a
verification problem. A verifier returns a verdict about an artifact, and every downstream
use of that verdict presumes we can say which artifact it was about. That means telling the
skill now loading from the one we checked, a first-party update from an adversarial fork,
and two names for one skill from two different skills.
Cryptographic hashing answers the narrowest version of this and, by design, no
more. Its avalanche property guarantees that a renamed variable decorrelates the digest, so
a hash cannot tell a refactor from a rewrite from an injection.
Verification without a rewrite-stable notion of identity therefore cannot accumulate: every
paraphrase is a new artifact, every verdict expires on contact with an edit, and nothing
links a tampered skill to the trusted base it was forked from.

No existing hashing scheme supplies it. A skill-identity signature must be
\emph{simultaneously} (a)~\textbf{rewrite-invariant}, so that a paraphrased prompt, a
renamed function, or a refactor still maps to the same skill;
(b)~\textbf{constant-size and $O(1)$-comparable}, so a registry can index millions of
skills; (c)~\textbf{component-localizing}, so that ``same code, different purpose'' is
distinguishable from ``same purpose, different code''; and (d)~\textbf{calibrated}, so the
score means something despite the compression. Cryptographic
hashing fails (a), (c), and (d) by design. Byte-level fuzzy hashes such as ssdeep and
TLSH~\citep{kornblum2006,oliver2013tlsh} target (a) but rely on spatial byte locality that
skills do not have; context-triggered piecewise hashing scored AUC $0.53$--$0.63$ on rewritten
skills. Raw embedding cosine is
accurate but stores kilobyte-scale float vectors, failing (b) and (d); plain
SimHash~\citep{charikar2002} gives a compact signature but was designed for single-document
web deduplication~\citep{manku2007}, failing (c) and (d). Per-channel clone detection for
skills~\citep{zhu2026skillreuse} supplies (c), but through dense vectors in a corpus-fit
basis, fused by a classifier trained on labeled pairs, so every party comparing scores must
share that basis and model.

We present a per-component, locality-sensitive fingerprint for agent skills that targets the
four-property conjunction, with (d) supplied by a prototype score normalization not yet
validated on held-out skills. We keep the philosophy of a fuzzy hash, a compact opaque
signature compared pairwise, but replace its engine. We decompose each skill into
components (prompt, code, tools), embed each with a learned encoder acting as a
\emph{convergent transformation} that maps diverse surface forms to nearby
vectors~\citep{reimers2019sbert,codebert2020}, and project each embedding to bits with a
multi-bank SimHash. The result is a signature of at most $120$ bytes whose components are
compared independently by Hamming distance and, deliberately, are never collapsed into a
single score until the application asks for one. A single similarity number tells you
\emph{whether} two skills are related; the per-component triple tells you \emph{where} they
differ, and which component moved. That is what deduplication, family clustering, novelty detection, and
provenance all need. Per-component similarity profiles for skills have precedent in clone
detection~\citep{zhu2026skillreuse}. \emph{What we contribute is a form of that profile a
verifier can build on}: a fixed-size bit code per component that needs no corpus-fitted
basis, codebook, or training labels and is compared at an a-priori threshold. We show that the triple is
readable and its components non-redundant. Whether it also discriminates better than one
whole-skill code of the same size is untested (Section~\ref{sec:limitations}).

We claim lineage, not behavioral equivalence, and Section~\ref{sec:twoaxis} measures that
boundary. Structural identity tells you that a skill is a tampered copy of a known one and
where it changed; whether the copy is safe is a separate question. What it buys a verifier is
routing. A verdict that attaches to a family survives the rewrites that invalidate a content
hash, and a match tells the verifier which verified base to diff against. Deciding where the
re-check starts then takes exact digests, because a fuzzy match cannot certify that a
component is unchanged (Section~\ref{sec:twoaxis}).

\paragraph{Findings.}
\begin{enumerate}[leftmargin=1.4em,itemsep=1.5pt,topsep=2pt]
  \item \textbf{The per-component triple shows which component moved.} Code is the strongest identity signal (AUC $0.997$) and prompt the weakest
    ($0.959$), and the clone label it supports flags only $0.01\%$ of pairs among real,
    independently authored skills. The $120$-byte triple tracks the embedding it
    approximates to within $0.004$ AUC per component and $0.03$ overall at $77\times$ fewer
    bits
    (Sections~\ref{sec:fidelity} and~\ref{sec:decomposition}).
  \item \textbf{Identity survives rewrites that erase surface text.} At each method's own
    zero-observed-false-positive threshold, on separately scored negatives, the fingerprint
    recovers all $50$ rewrites across ten transforms and a lexical baseline $45$, the
    difference being five full rewrites; at its a-priori threshold the fingerprint recovers
    $49$--$50$ (Section~\ref{sec:robustness}).
  \item \textbf{Structural identity complements behavioral verification, and we locate the
    seam.} On a $906$-skill injection benchmark the fingerprint recognizes $93\%$ of generated
    injections as tampered copies of a registered base and localizes the change mostly to the
    edited component, but as a safety detector reaches at best $F_1$ $0.735$ against behavioral
    verification's $0.946$. Recognition requires the base to be registered: only $3$ of $44$
    in-the-wild malicious skills match a benign skill (Section~\ref{sec:twoaxis}).
\end{enumerate}

%% file: sections/method.tex
\section{The per-component fingerprint}
\label{sec:method}

The fingerprint is defined in three steps: decompose a skill into components, embed and
project each component to bits independently, and compare two fingerprints component by
component (Figure~\ref{fig:pipeline}). The design choice the paper examines is to keep the
components separate instead of hashing a skill as one blob.

\begin{figure}[t]
  \centering
  \resizebox{0.58\linewidth}{!}{%
  \begin{tikzpicture}[
    font=\footnotesize,
    comp/.style={rounded corners=2pt, draw=#1, fill=#1!12, text=#1,
                 minimum width=1.7cm, minimum height=0.85cm, align=center, thick},
    stage/.style={rounded corners=2pt, draw=gray!60, fill=white,
                  minimum width=1.7cm, minimum height=0.85cm, align=center},
    codeb/.style={rounded corners=2pt, draw=#1, fill=#1!12, text=#1,
                  minimum width=1.0cm, minimum height=0.85cm, align=center},
    hdr/.style={align=center, font=\footnotesize\bfseries},
    arr/.style={-{Stealth[length=2mm]}, draw=gray!65, semithick},
  ]
    \node[hdr] at (0,1.05)   {skill\\components};
    \node[hdr] at (2.7,1.05) {embed\\(768-d)};
    \node[hdr] at (5.4,1.05) {Multi-SimHash\\$5\times64$};
    \node[hdr] at (7.9,1.05) {code\\(40\,B)};

    \node[comp=cprompt] (p0) at (0,0)     {prompt};
    \node[comp=ccode]   (c0) at (0,-1.35) {code};
    \node[comp=ctools]  (t0) at (0,-2.7)  {tools};
    \foreach \r/\y in {p/0, c/-1.35, t/-2.7} {
      \node[stage] (\r1) at (2.7,\y) {$E(\cdot)$};
      \node[stage] (\r2) at (5.4,\y) {sign$(Wv)$};
    }
    \node[codeb=cprompt] (p3) at (7.9,0)     {40\,B};
    \node[codeb=ccode]   (c3) at (7.9,-1.35) {40\,B};
    \node[codeb=ctools]  (t3) at (7.9,-2.7)  {40\,B};

    \node[rounded corners=3pt, draw=cfp, fill=cfp!10, text=cfp, thick,
          minimum width=1.7cm, minimum height=3.7cm, align=center]
          (fp) at (10.1,-1.35) {120-byte\\fingerprint\\(triple)};

    \foreach \r in {p,c,t} {
      \draw[arr] (\r0) -- (\r1);
      \draw[arr] (\r1) -- (\r2);
      \draw[arr] (\r2) -- (\r3);
      \draw[arr] (\r3.east) -- (fp.west);
    }

    \node[align=center, font=\footnotesize\itshape] at (4.9,-3.75)
      {compare: per-component
       $\mathrm{sim}=1-\mathrm{popcount}(F_A\oplus F_B)/320$};
  \end{tikzpicture}}
  \caption{The per-component fingerprint pipeline. Each component is embedded
  independently, projected to a $320$-bit multi-bank SimHash code, and the codes form a
  triple of at most $120$ bytes compared component-wise by Hamming distance. The triple is never
  collapsed to a scalar inside the fingerprint itself.}
  \label{fig:pipeline}
\end{figure}

\paragraph{The convergent transformation.}
We treat a skill as a small structured artifact $S = \{c_1, \dots, c_m\}$ whose
\emph{components} are textual and drawn from a fixed vocabulary: the natural-language
\emph{prompt}, the \emph{code} that implements the skill, and the \emph{tool} declarations
it uses. Because skills lack the spatial byte locality a fuzzy hash exploits, we require a
\emph{convergent transformation} $f$ under which functional equivalence of components implies
geometric proximity of outputs. A learned text encoder approximates such an $f$ for the
rewrite families we test: we embed each component with $E(\cdot) \to \mathbb{R}^{d}$
(text-embedding-005, $d = 768$, in all main experiments), which maps paraphrases, renamings,
and refactorings to nearby vectors~\citep{reimers2019sbert,codebert2020,wang2022e5}. The
engine's default configuration normalizes Python code on its abstract syntax tree before
embedding (identifiers alpha-renamed, comments and docstrings stripped, raw text kept for code
that does not parse); this is recorded for the adversarial benchmark and the re-embeddings of
Appendices~\ref{app:quant} and~\ref{app:ablation} (see Section~\ref{sec:fidelity}).
Everything downstream inherits the encoder's quality. The method is encoder-agnostic
(Appendix~\ref{app:ablation} ablates four), but codes are comparable only under the same
encoder, preprocessing, projection seeds, and bit layout, so a protocol tag recording them
travels with every signature.

\paragraph{Multi-bank SimHash.}
Storing a $768$-dimensional float vector per component defeats the purpose of a compact,
indexable signature. We compress each embedding to bits with SimHash~\citep{charikar2002},
taking one bit per random Gaussian hyperplane $r \sim \mathcal{N}(0, I_d)$ as
$b(v; r) = \mathbb{1}[\, v \cdot r \ge 0 \,]$. We draw $B = 320$ independent hyperplanes per
component and pack the bits into five $64$-bit words, the ``banks'' that name the method.
The five banks are a machine-word layout over one $320$-bit code: all bits are i.i.d., so the
packing buys word-aligned comparison without changing accuracy. The projection matrix is
seeded deterministically per component, so codes from different components are never
compared. A skill's signature holds one $320$-bit code per present component,
\textbf{$40$ bytes each, a payload of at most $120$ bytes} plus the protocol tag, regardless of
skill size. For a component
present in both skills, with codes $F_A$ and $F_B$, similarity is the complement of the
normalized Hamming distance,
\begin{equation}
  \mathrm{sim}(F_A, F_B) \;=\; 1 - \frac{\mathrm{popcount}(F_A \oplus F_B)}{B},
  \qquad B = 320,
  \label{eq:hamming-sim}
\end{equation}
which evaluates in under a microsecond. Where an application needs one number, the
\emph{overall similarity} is a weighted mean over the set $P$ of components present in either
skill,
\begin{equation}
  \mathrm{sim}(A, B) \;=\; \frac{\sum_{i \in P} w_i\, s_i}{\sum_{i \in P} w_i},
  \qquad (w_{\text{prompt}}, w_{\text{code}}, w_{\text{tools}}) = (0.30, 0.35, 0.20),
  \label{eq:overall}
\end{equation}
where $s_i$ is Equation~\ref{eq:hamming-sim} for a component present in both skills and $0$
for one present in only one, so a missing component lowers the score; a component absent from
both is skipped. The weights are set by design, code highest as the most behavior-proximal
component, and are not fit to any evaluation corpus. Every overall score, overall AUC, and the
$0.85$ operating threshold in this paper refer to Equation~\ref{eq:overall}. The penalty is
deliberate and has a cost: a copy with its tools removed scores $0.65/0.85 = 0.765$, below the
threshold, so component removal is surfaced through the per-component pattern for review
instead of as a near-clone. Two quantities are reported alongside but enter neither score nor
label: \emph{coverage}, the Jaccard overlap of present components, and \emph{asymmetry}, the
spread between the most and least similar components.

\subsection{Why the bits preserve identity}
\label{sec:theory}

Compressing a $768$-dimensional embedding to $320$ bits preserves the ordering of similar and
dissimilar skills, at a quantifiable cost. The expectation of Equation~\ref{eq:hamming-sim}
is $\mathbb{E}[\mathrm{sim}] = 1 - \arccos(c)/\pi$ at $c = \cos(u,v)$, strictly increasing in
$c$ and hence an order-preserving bijection~\citep{charikar2002}. Because AUC depends only on
score ranking, a component's SimHash AUC converges to its cosine AUC as bits grow. The
weighted overall score of Equation~\ref{eq:overall} has no such guarantee, since a weighted
mean of transformed cosines need not rank pairs like the weighted mean of cosines; its
fidelity is an empirical finding (Appendix~\ref{app:quant}). The pairing is
principled: a learned encoder is compared by cosine, and SimHash is the locality-sensitive
family whose collision probability is a function of that same angle. We discard bits, not the
metric. At $B = 320$ the estimator's standard deviation is $\sigma \approx 0.023$, and a
Hoeffding bound turns that average into a per-comparison guarantee holding with high
probability (Appendix~\ref{app:bitbudget}). Those $320$ bits are $\mathbf{77\times}$ smaller
than the $24{,}576$-bit float embedding.

\paragraph{The negative floor and the operating threshold.}
Two effects set the score scale, and they are easily conflated (Figure~\ref{fig:theory}).
First, unrelated skill components are \emph{not} orthogonal in embedding space, because text
and code share vocabulary and learned encoders are anisotropic~\citep{li2020anisotropy}, so
unrelated pairs carry a positive background cosine (mean $0.54$ in our corpus). That
baseline lifts the negative floor: $c = 0.54$ maps to $\approx 0.68$, where
orthogonal vectors would sit at
$0.5$. Second, the transfer function's slope, $1/(\pi\sqrt{1-c^2})$, is only $0.38$--$0.51$
over the observed cosine range, so it \emph{compresses} the gap on top of that floor, from
$0.236$ in cosine space to an observed $0.109$ at $320$ bits. The useful discrimination range
therefore compresses into roughly $[0.65, 1.0]$. Every identity threshold in this paper,
including the $0.85$ operating point, is set from this floor, observed on the constructed
corpus, and then held fixed on every other corpus (Appendix~\ref{app:registry}). The one tuned threshold is the safety-detector baseline of
Section~\ref{sec:twoaxis}, reported at its in-sample optimum as an upper bound. Theory bounds what the compression costs. Whether the
embedding separates skills in the first place is an empirical question, measured next.

%% file: sections/results.tex
\subsection{Compression fidelity}
\label{sec:fidelity}

The $120$-byte signature tracked the ranking of the $24{,}576$-bit embedding to within
$\mathbf{0.004}$ AUC per component and $0.03$ overall (Appendix~\ref{app:quant}). We measured
fidelity on a corpus built to isolate each rewrite axis:
$100$ skills across $40$ domains as $20$ groups of $5$, each holding an
\emph{original} and four controlled rewrites, a \emph{clone}
(prompt rewritten), a \emph{repurposed} skill (same code, new domain), a
\emph{reimplemented} skill (same prompt, new code), and a \emph{toolswap} (tools removed).
That gives $4{,}950$ comparisons ($200$ positive, $4{,}750$ negative), scored by AUC, which
measures ranking discrimination. Overall AUC was $\mathbf{0.974}$
(95\% CI $[0.956, 0.994]$ from a skill-level bootstrap accounting for shared-skill
dependence among the pairs), with a positive mean of $0.876$ against a negative mean of
$0.623$. That figure, with Tables~\ref{tab:percomp} and~\ref{tab:patterns} and the family
results of Section~\ref{sec:decomposition}, is from the original run, whose code preprocessing
was not recorded. For the default configuration of Section~\ref{sec:method}, the reproducible
figure is $0.931$, from a re-embedding with AST-normalized code, against $0.959$ for cosine.
The difference is confined to code: prompt and tools mean similarities agree within $0.005$,
while the code negative mean rose from $0.677$ to $0.823$, consistent with alpha-renaming
pulling unrelated code together. The bootstrap interval does not capture this dependence on
preprocessing.

A method ladder on aggressive single-component rewrites separates the encoder's contribution
from the hash's (Table~\ref{tab:ladder}). Byte-level approaches failed: context-triggered
piecewise hashing scored AUC $0.53$--$0.61$, and hashing a normalized abstract syntax tree
(AST) reached only $0.63$, recovering exact (Type-2) clones but collapsing on refactors
(Type-3). An LLM that rewrote each component into a constrained vocabulary before hashing
did better ($0.89$) at the cost of latency and model noise. Embedding cosine was the ceiling
($0.99{+}$), and the multi-bank SimHash matched it ($0.992$--$0.996$) at $77\times$ less
storage. Invariance therefore comes from the embedding stage (the encoder and, where used,
code normalization) and compactness from the hash, and among compressions of comparable
accuracy SimHash is the only one that needs no corpus-fitted basis or codebook and is
Hamming-indexable (Appendix~\ref{app:quant}). Because its variants follow the same
invariance axes the embedding absorbs, this corpus establishes fidelity and nothing harder.
Several AUCs appear across this paper under different corpora, which
Appendix~\ref{app:metricmap} keeps distinct. We now stress the fingerprint where identity is
hardest to preserve.

%% file: sections/robustness.tex
\section{Adversarial robustness of the identity signal}
\label{sec:robustness}

An identity signal is useful to a verifier only if rewriting does not escape it. Against ten
transforms up to cross-language translation, the fingerprint recovered all $\mathbf{50}$
rewrites with zero observed false positives in every run, including five full rewrites of
both components that a lexical baseline missed even at its own zero-observed-false-positive
threshold.

\paragraph{The rewrite benchmark.}
We built $50$ positive pairs from five source skills, each transformed by ten rewrite
operators: prompt paraphrase, restyling, and obfuscation; code renaming, refactoring, and
minification; cross-language translation to JavaScript, TypeScript, and Go; and a full
rewrite of both components. All rewrites are LLM-generated and not functionally tested, so they
establish similarity to generated descendants and leave semantic equivalence unverified, and
none is optimized against the fingerprint (Section~\ref{sec:limitations}). These transforms are
the worst case for a surface signature, because a translated component shares almost no
character $n$-grams with its original. In the run reported here, overall similarity stayed
above $0.85$ on every transform (minimum $0.859$, on a full rewrite), and the per-component
triple shows why: an
attack on one component left the others at $1.0$. Prompt attacks dropped prompt similarity to
$0.74$--$0.88$ while code stayed at $1.000$, and conversely. Cross-language translation was
the hardest code transform (code ${\approx}\,0.82$). A lexical TF-IDF baseline collapsed on
exactly these transforms, falling to $0.14$ for paraphrase, $0.20$ for obfuscation, $0.38$
for minification, and ${\approx}\,0.44$ for translation, against the fingerprint's
$0.74$--$0.98$ (Figure~\ref{fig:adversarial} plots every transform).

\paragraph{Operating points: ranking AUC, shared thresholds, and false-positive evidence.}
A fair comparison of the two methods needs more care than either ranking AUC or a shared
threshold provides. Ranking AUC was near $1.0$ for \emph{both} methods, because the negatives
are cross-source skills with almost no surface overlap (TF-IDF negative mean $0.07$), so AUC
rewards separating easy negatives even as positive scores collapse. A shared threshold
misleads in the other direction. At $0.85$, TF-IDF recovered only $5/50$, but that threshold
comes from the fingerprint's negative floor (Section~\ref{sec:theory}) and has no meaning on
TF-IDF's scale. At its own zero-false-positive threshold ($0.35$, the maximum over its
$1{,}210$ cross-source negatives), TF-IDF recovered $45/50$ (Table~\ref{tab:adv}). Comparing
each method at its own zero-observed-false-positive point shows where they differ, though the
two zeros are unequally strong: $0/10$ negatives bounds the fingerprint's false-positive rate
only below $26\%$ ($95\%$ one-sided), while $0/1{,}210$ bounds TF-IDF's below $0.25\%$. Both
methods recovered every single-component rewrite. They split on the five full rewrites of both components, which the fingerprint
recovered and TF-IDF scored at $0.16$--$0.29$, below its own negative maximum. The
fingerprint also fixed its threshold before seeing this benchmark, whereas TF-IDF's
zero-false-positive point was read off these negatives after the fact. At that a-priori
threshold the count is less stable: the closest full rewrite cleared $0.85$ by $0.009$, under
one standard deviation of the estimator, and a repeat run recovered $49/50$. The benchmark
is small, so we report it as a stress test of transform-invariance. A powered
comparison at matched false-positive rates on held-out, overlapping negatives remains owed.
Cross-language is the boundary: controlled translation is recovered because the unchanged
prompt carries it, while an independently authored reimplementation in another language,
which would share no component, was not evaluated and is expected to fail
(Appendix~\ref{app:crosslang}).

\begin{table}[t]
  \centering
  \small
  \caption{Adversarial benchmark ($50$ positive pairs, $10$ transforms) and the constructed
  corpus. Ranking AUC is near-saturated for both methods. The shared $0.85$ threshold is the
  fingerprint's a-priori operating point and understates TF-IDF; at each method's
  zero-observed-FP point they differ only on the five full rewrites, on unequal negative
  sets. Fingerprint rows span
  two runs of the default configuration ($10$ cross-source negatives each); embedding raw
  code without AST normalization gives $45/50$ at $0.85$ and $50/50$ at $0.80$. TF-IDF is
  scored on $1{,}210$ negatives. On the constructed corpus a lexical baseline
  \emph{matches} the fingerprint. $^{\dagger}$Original configuration; $0.931$ under the
  default AST configuration (Section~\ref{sec:fidelity}).}
  \label{tab:adv}
  \setlength{\tabcolsep}{4.5pt}
  \begin{tabular}{lcccccc}
    \toprule
    & \multicolumn{4}{c}{adversarial rewrites} & \multicolumn{2}{c}{constructed corpus} \\
    \cmidrule(lr){2-5}\cmidrule(lr){6-7}
    Method & AUC & threshold & det. & FP & AUC & bytes/skill \\
    \midrule
    Char-$n$-gram TF-IDF & $0.998$ & $0.85$ (shared) & $5/50$ & $0$
      & $\mathbf{0.999}$ & ${\approx}8{,}400$ (sparse) \\
    \quad own zero-FP point & & $0.35$ (post hoc) & $45/50$ & $0$ & & \\
    Fingerprint (ours) & $1.000$ & $0.85$ (a priori) & $49$--$50/50$ & $0$
      & $0.974^{\dagger}$ & $\mathbf{{\le}\,120}$ \\
    \quad own zero-FP point & & ${\le}\,0.80$ (post hoc) & $\mathbf{50/50}$ & $0$ & & \\
    \bottomrule
  \end{tabular}
\end{table}

The same surface overlap explains why the lexical baseline \emph{matches} the fingerprint on
the fidelity corpus of Section~\ref{sec:fidelity} (AUC $0.999$ versus $0.974$): every designed
variant there keeps one component byte-identical and the negatives are cross-domain, so
shared surface vocabulary separates them. The fingerprint's standing advantage is a
bounded $120$-byte signature compared in $O(1)$ by Hamming distance, against a TF-IDF
vector averaging ${\approx}8.4$\,kB per skill that is not bit-comparable, a $70\times$ gap
that decides feasibility at scale.

\paragraph{Label quality on a community corpus.}
Verification signals for agentic tasks are often described as noisy, and on real skills part
of that noise sat in the ground truth. On $1{,}000$ community skills the per-component AUCs
stayed nearly unchanged from their constructed values, while the overall AUC fell to $0.739$
against name-based duplicate labels. Six of the $13$ positive pairs share only a name and
scored $0.27$--$0.53$. Inspection suggests some of these labels conflate distinct skills: two
independently written skills both named \texttt{prompt-optimizer} scored $0.265$, one doing
iterative refinement and the other single-pass style transfer. This reading comes from
inspecting pairs the fingerprint itself scored low and rests on $13$ positives, so we report
it as an observation about label quality that independent human annotation would need to
confirm (Appendix~\ref{app:community}). Surviving rewrites is what an identity signal owes a
verifier. What it is worth depends on what can be read off it.

%% file: sections/decomposition.tex
\section{Beyond a scalar: component profiles, families, and novelty}
\label{sec:decomposition}

A single similarity number tells you \emph{whether} two skills are related. The triple tells
you \emph{where} they agree and differ, which is the difference between a score and an
identity primitive. What makes that readable is that the components carry sharply different
signal. On the corpus of
Section~\ref{sec:fidelity}, \emph{code} is the strongest discriminator (AUC $\mathbf{0.997}$)
and \emph{prompt} the weakest yet still strong ($0.959$), as Table~\ref{tab:percomp} reports.
\emph{Tools} reach $1.000$, but only on the $7.6\%$ of pairs where both skills declare tools,
a subset this corpus makes trivially separable. The ordering is the empirical content of the
design principle that code is the most behavior-proximal component: intent paraphrases
freely while implementation does not.
Averaging that spread away is exactly what a scalar score does. The components are
also non-redundant. Among same-group pairs the per-component similarities are uncorrelated
to moderately anticorrelated (prompt versus code $r = -0.40$), and across the diverse SkillsBench
corpus~\citep{li2026skillsbench} they are nearly independent ($r = 0.08$ over $26{,}565$
pairs). That independence is what lets repurposing (high code, low prompt) and reimplementation
(high prompt, low code) leave different per-component patterns. These properties show that the triple is readable
and non-redundant. They do not show that it discriminates better than one code of the same
size computed over the whole skill, which we did not test (Section~\ref{sec:limitations}). In
a supervised lexical setting, SkillReuse's ablation found that per-channel features added
$6.1$ $F_1$ points over flat TF-IDF~\citep{zhu2026skillreuse}, which is suggestive for our
setting without settling it.

\begin{table}[t]
  \centering
  \small
  \begin{minipage}[t]{0.46\linewidth}
    \centering
    \caption{Per-component discrimination on $4{,}950$ pairs. Coverage differs by component:
    $28\%$ of skills declare tools, so both skills have them in only $7.6\%$ of pairs and
    that AUC rests on a subset. Original-configuration values (Section~\ref{sec:fidelity}).}
    \label{tab:percomp}
    \setlength{\tabcolsep}{3.6pt}
    \begin{tabular}{lcccc}
      \toprule
      Component & Pos. & Neg. & Gap & AUC \\
      \midrule
      prompt        & $0.867$ & $0.704$ & $0.163$ & $0.959$ \\
      \textbf{code} & $0.930$ & $0.677$ & $0.253$ & $\mathbf{0.997}$ \\
      \textbf{tools}& $1.000$ & $0.758$ & $0.242$ & $\mathbf{1.000}$ \\
      \midrule
      overall       & $0.876$ & $0.623$ & $0.254$ & $0.974$ \\
      \bottomrule
    \end{tabular}
  \end{minipage}\hfill
  \begin{minipage}[t]{0.50\linewidth}
    \centering
    \caption{Relationship classification. The triple recovers reuse within groups while
    almost never flagging false clones across groups.}
    \label{tab:patterns}
    \begin{tabular}{llc}
      \toprule
      Pair set & Top relationship & Share \\
      \midrule
      same group      & clone      & $\mathbf{62.0\%}$ \\
      same group      & tool\_swap & $24.0\%$ \\
      same group      & variant    & $14.0\%$ \\
      \midrule
      different group & variant    & $98.2\%$ \\
      different group & clone (false pos.) & $\mathbf{0.2\%}$ \\
      \bottomrule
    \end{tabular}
  \end{minipage}
\end{table}

\paragraph{Relationship labels: a coarse heuristic over an informative profile.}
Because the components are kept separate, the pattern of high and low per-component
similarities is visible with no extra model. The prototype maps it to six labels (clone,
reimplemented, repurposed, tool swap, variant, unrelated, with a mixed fallback) by fixed
rules in priority order over the three component scores, a missing component scoring $0$,
with cutoffs taken from the SimHash floor of Section~\ref{sec:theory}
(Appendix~\ref{app:registry}). The profile is more informative than the labels. A CSV analyzer
and a health-metrics skill share code similarity $1.000$ but prompt similarity $0.772$, the
signature of repurposing that a scalar hides; the rules, which reserve repurposed for prompts
below $0.60$, report a tool swap because neither skill declares tools. Because the SimHash floor
compresses the low end, a rewritten component rarely falls below the high cutoff of $0.80$:
in a controlled test a reimplementation (prompt $1.000$, code $0.831$) was labeled a clone.

Against group membership the labels are sharply asymmetric. The triple assigned a reuse
label to $86\%$ of the $200$ same-group pairs, clone for $124$ and tool-swap for $48$ (a
label that, as coded, also covers pairs without tools; Appendix~\ref{app:registry}), while
calling only $0.2\%$ of the $4{,}750$ cross-group pairs clones (Table~\ref{tab:patterns}); the
variant label, given to $98.2\%$ of cross-group pairs, carries little information.
Against the designed variant types they are coarse: no same-group pair was labeled repurposed
or reimplemented, so at these cutoffs the labels recover reuse at the group level while the
per-component scores carry the finer pattern. Resolving the finer types would need cutoffs
set on the normalized scale of Appendix~\ref{app:registry}, which we have not validated. The
production cost of a clone detector, though, is flagging genuinely different but
\emph{overlapping} real skills, which a corpus with cross-domain negatives cannot measure.
On the $502$ real, independently authored benign skills of the injection benchmark
(Section~\ref{sec:twoaxis}), across the $125{,}678$ pairs with distinct names (excluding $73$
same-name pairs, mostly one skill present in two source collections), the clone-flag rate at
the operating threshold was $\mathbf{0.01\%}$ ($11/125{,}678$; $0.10\%$ at the looser
$0.80$). Inspection showed all eleven to be genuine reuse families, each attributed by the
triple to the component the two skills share, so the false-clone rate is lower still
(Appendix~\ref{app:registry}).

\paragraph{Novelty routing at registry scale.}
The same distances reduce to the triage operation a verifier needs at scale. Density-based
clustering~\citep{campello2013hdbscan,mcinnes2017hdbscan} over the per-component Hamming
distances groups skills into \emph{families} without specifying the cluster count in advance,
and a skill matching no family is flagged \emph{novel} and routed to review. The fingerprint
recovered this structure cleanly on the $20$-group ground truth: a skill's nearest neighbor
was a true same-group variant for $96\%$ of skills, clustering at $0.80$ produced $23$
families at $90\%$ purity, and all $100$ held-out skills from withheld families were flagged
novel at the operating threshold with zero mis-merges. Each $40$-byte component code is
exactly what a binary Hamming index consumes~\citep{johnson2019faiss,manku2007}, putting a
million skills at $40$\,MB per component. Families, canonical members, novelty, and a
portable \emph{SkillBOM}~\citep{cyclonedx} designed to carry the fuzzy fingerprint next to
per-component cryptographic digests are all reductions of the same triple
(Appendix~\ref{app:registry}). Every use so far treats a high-similarity match as good news.
The next section examines the case where it is not.

%% file: sections/twoaxis.tex
\section{Identity and integrity: two composing axes}
\label{sec:twoaxis}

An injected malicious skill is, structurally, the benign skill it was injected into. That
makes it the cleanest test of what a structural fingerprint can and cannot do. It should
recognize the injected skill as a tampered copy of a known base, and it should \emph{not},
on its own, decide whether the skill is safe. Neither role requires resisting an attacker
who optimizes against the fingerprint (Section~\ref{sec:limitations}). We test both on the
$906$-skill benchmark of our companion behavioral integrity verification (BIV)
work~\citep{wu2026biv}: $502$ benign skills, $44$ in-the-wild malicious
skills~\citep{liu2026maliciousskills}, and $360$ injections that two
generators~\citep{schmotz2026skillinject,jia2026skillject} built from clean base skills, many
of them also in the benign pool, in the indirect-prompt-injection
tradition~\citep{greshake2023injection,zhan2024injecagent}.

\paragraph{Recognition and localization of injected tampering.}
The fingerprint recognized most injected skills as copies of a known base when that base
was registered (Table~\ref{tab:biv_2x2}). Of the $360$ generated injections, $\mathbf{335}$
($93\%$) were near-clones (similarity $\ge 0.85$) of a benign skill: $179/200$ for skillject
and $156/160$ for skill-inject. Of the $44$ in-the-wild malicious skills, only $\mathbf{3}$
were. The pooled $83.7\%$ ($338/404$) therefore measures the threat model the generators
construct, tampering with a skill whose clean version is known, and not how common that case
is in the wild. Within that threat model the injection does not hide the lineage; it rides on
top of it. The match is also the right one: $237$ of the $335$ matches are the generator's
clean copy of the base, and $98$ a same-named (or, twice, near-named) skill from another
collection whose instruction lines reappear verbatim in the injection, at least $90\%$ of
them in $90$ cases (median $98\%$). Lineage is thus directly evidenced for $327$ matches; the
other $8$ rest on name and fingerprint.

\begin{figure}[t]
  \centering
  \includegraphics[width=0.56\linewidth]{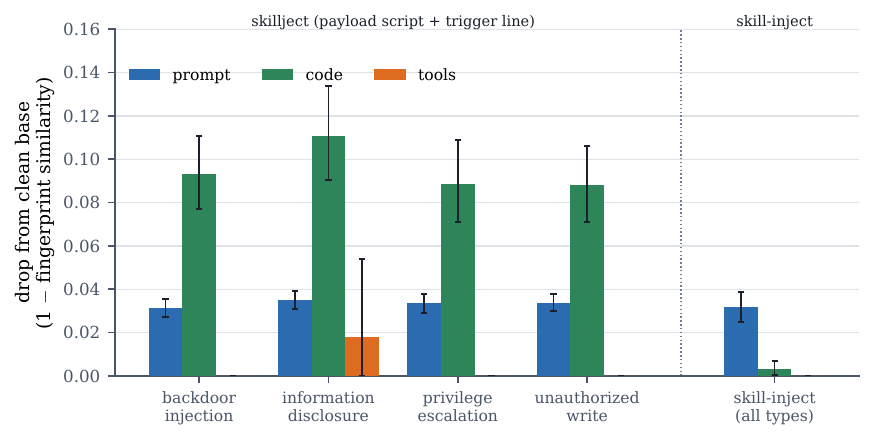}
  \caption{Tamper localization in fingerprint space. Per-component drop from the clean base
  ($1-$ fingerprint similarity) for each injection paired with its clean base; bars are means
  with $95\%$ bootstrap intervals. Skillject ($46$--$50$ pairs per attack type) adds a
  payload script and a trigger line: code drops most (similarity $0.89$--$0.91$) and the
  prompt less ($0.967$). Skill-inject ($62$ paired injections) edits mainly the instructions:
  the prompt drops while code stays at $0.997$. Tools are present in both skills for only $8$
  pairs per skillject attack type and $2$ skill-inject pairs.}
  \label{fig:biv_tamper}
\end{figure}

The per-component triple then localized the larger change, pair by pair, to the component
each generator mainly edits (Figure~\ref{fig:biv_tamper}). We paired each injection with its clean base and compared
fingerprints. Skillject adds a payload script and a line instructing the agent to run it.
Across its four attack types, code similarity fell to $0.89$--$0.91$ and prompt similarity to
about $0.967$, and code moved more than the prompt in $159$ of $184$ pairs. Skill-inject
edits mainly the instruction text ($52$ of its $62$ paired injections change the prompt). Code stayed at $0.997$ while the prompt fell to $0.968$, and
the prompt moved more in $41$ of $51$ pairs. With no attack-specific tuning, the fingerprint
reports ``a modified copy of base $X$, changed mostly in the code'' for one generator and
``mostly in the prompt'' for the other. The in-the-wild skills have no clean base to
localize against.

\paragraph{The limit of the identity axis.}
The same fact that makes the fingerprint a good lineage signal makes it a poor safety
detector: when malicious skills are near-clones of benign ones, structural similarity cannot
separate them. Used as a detector (flagging a skill when its fingerprint matches a
known-malicious skill of a different base, leave-base-out), it reached $F_1$ $0.735$ at a
$32\%$ false-positive rate, and that only at its $F_1$-optimal threshold ($0.781$), chosen on
the evaluation set and therefore an upper bound. At the fixed $0.85$ operating point it
reached $F_1$ $0.217$ (Table~\ref{tab:biv_detect}). It beat a rule-based scanner ($0.44$) by
recognizing repeated injection payloads as a family, but fell well short of both behavioral
methods on the same benchmark, an LLM-only audit ($0.93$) and BIV
($0.946$)~\citep{wu2026biv}. That gap marks the boundary of the identity axis: the
fingerprint answers ``is this a copy of a skill we know?'', not ``does this skill do what it
claims?''

\begin{table}[t]
  \centering
  \small
  \begin{minipage}[t]{0.50\linewidth}
    \centering
    \caption{Malicious-skill detection on the $906$-skill benchmark. Structural similarity
    recognizes injection-payload families but cannot match behavioral verification. The
    fingerprint's best row uses an in-sample threshold. Behavioral and baseline numbers are
    from~\citet{wu2026biv}.}
    \label{tab:biv_detect}
    \setlength{\tabcolsep}{3.5pt}
    \begin{tabular}{lcccc}
      \toprule
      Method & axis & $F_1$ & rec. & FPR \\
      \midrule
      Rule-based scanner & n/a & $0.44$ & n/a & n/a \\
      Fingerprint, $t{=}0.85$ & ident. & $0.217$ & $0.13$ & $0.03$ \\
      Fingerprint, best $t$ & ident. & $0.735$ & $0.81$ & $0.32$ \\
      LLM-only audit & behav. & $0.93$ & n/a & n/a \\
      Behavioral (BIV) & behav. & $\mathbf{0.946}$ & $0.978$ & $0.072$ \\
      \bottomrule
    \end{tabular}
  \end{minipage}\hfill
  \begin{minipage}[t]{0.47\linewidth}
    \centering
    \caption{Structural familiarity (near-clone of a benign skill, $\ge 0.85$) versus the
    behavioral label, malicious skills split by source. The $335$ generated \emph{tampered
    known} skills need both axes. Of the $37$ familiar benign skills, $24$ match only a
    same-name copy from another collection and $13$ form the $11$ reuse pairs of
    Section~\ref{sec:decomposition}.}
    \label{tab:biv_2x2}
    \setlength{\tabcolsep}{3.5pt}
    \begin{tabular}{lccc}
      \toprule
      & & \multicolumn{2}{c}{malicious} \\
      \cmidrule(lr){3-4}
      & benign & generated & in the wild \\
      \midrule
      near benign & $37$ & $\mathbf{335}$ & $3$ \\
      novel               & $465$ & $25$ & $41$ \\
      \bottomrule
    \end{tabular}
  \end{minipage}
\end{table}

\paragraph{A runtime admission policy from the two axes.}
The axes are complementary, and the generated tampered-known cell shows why
(Table~\ref{tab:biv_2x2}). Such a skill is a near-clone of a benign skill, so the identity
axis says ``known, trusted-looking.'' It also carries an injected capability, so the
behavioral axis says ``malicious.'' Neither axis handles it alone. A structural registry
waves it through as known, and a behavioral check that ignores lineage cannot tell a
first-party update from an adversarial fork. The same failure mode covers benign change: a
modification the identity axis admits passes silently whether an adversary or the skill's
own author introduced it, which is what makes a regression hard to catch.

The composition suggests a policy for admitting skills at runtime, with one rule that keeps
similarity from standing in for trust: a fuzzy match routes and never certifies. An arrival
with no family match goes to a full behavioral audit. An arrival that matches a registered
base is diffed against that base, and its behavioral check starts from the components whose
cryptographic digest differs from the verified base. The check must also cover how those
components interact with the unchanged ones, since an edited prompt can drive unchanged code
in new ways, and authorship needs separate provenance such as a signature. The benchmark shows
why the starting point must be exact.
Of the $262$ paired injections, $252$ edited the prompt, and every one of those edited prompts
still scored above $0.85$ (minimum $0.872$). For skillject that edit is the line that tells
the agent to run the payload, so a check scoped by fuzzy similarity would skip the very line
that triggers the attack. The policy has a precondition, which the in-the-wild result makes
plain: it helps only when the clean base is already registered, as when a verified skill
returns as an update, fork, or re-upload. A content hash alone cannot route at all, because
every one of these injections changes bytes and arrives as a new artifact. Any behavioral
checker can sit on the other axis; the LLM-only audit nearly matches BIV here and would be
routed the same way, and we pair with BIV because its typed, per-capability diff aligns with
per-component localization. We measured neither the combination nor the verification cost
the cascade saves, so we offer it as a policy this result licenses.

When agents write skills, near-duplicates multiply faster than any audit budget, and
family-level identity keeps oversight tractable: it tells a verifier which earlier verdict is
relevant and records what moved since the last check. Safety itself stays with
environment-grounded behavioral checking.

%% file: sections/related.tex
\section{Related work}
\label{sec:related}

\paragraph{Hashing, embeddings, and clones.}
The fingerprint is built on SimHash~\citep{charikar2002}, the signed-random-projection
member of the locality-sensitive hashing
family~\citep{indyk1998,gionis1999,datar2004,broder1997}; its angular collision probability,
with a Hoeffding bound, is why so few bits preserve rankings, as
Johnson--Lindenstrauss~\citep{jl1984} does for real-valued projections. Context-triggered piecewise hashing (ssdeep)~\citep{kornblum2006} and
TLSH~\citep{oliver2013tlsh} inspired our framing, but they exploit spatial byte locality that
skills lack, so we replace their byte engine with a learned
encoder~\citep{reimers2019sbert,wang2022e5,muennighoff2022mteb} that captures program
semantics~\citep{codebert2020,guo2021graphcodebert,husain2019codesearchnet}. Our rewrite axes
mirror the clone taxonomy~\citep{roy2009clone} used by token- and tree-based
detectors~\citep{sajnani2016sourcerercc,jiang2007deckard}.

\paragraph{Per-channel skill clone detection.}
SkillReuse~\citep{zhu2026skillreuse}, first released as SkillClone, is the closest prior
work. It splits each \texttt{SKILL.md} into YAML, instructions, and code, encodes each channel
with corpus-fit TF-IDF and SVD, and fuses the per-channel similarities, their cross-channel
products, structure features, and a flat TF-IDF score with a logistic model trained on labeled
pairs, reaching $F_1$ $0.939$ against $0.881$ for flat TF-IDF on $300$ pairs (removing the
channel features costs $6.1$ points). Per-channel profiles are therefore established; ours
differs as a $40$-byte, Hamming-indexable code per component with no corpus-fitted basis,
codebook, or labels, an a-priori threshold, tools as a component, and use as an identity
primitive alongside behavioral verification.

\paragraph{Agents, attacks, and provenance.}
Tool-using agents~\citep{yao2022react,schick2023toolformer,patil2023gorilla,qin2023toolllm}
make skill identity necessary, and the Model Context Protocol~\citep{mcp} and agent skill
formats~\citep{anthropicskills} are the artifacts our fingerprint spans. The
threat behind the two-axis framing is studied as indirect prompt
injection~\citep{greshake2023injection} and malicious tool use~\citep{zhan2024injecagent},
evaluated in emulated sandboxes~\citep{ruan2023toolemu} and attack-and-defense
environments~\citep{debenedetti2024agentdojo}; those target runtime behavior, the
complementary axis (Section~\ref{sec:twoaxis}). The SkillBOM extends bill-of-materials
standards~\citep{cyclonedx}, model cards and
datasheets~\citep{mitchell2019modelcards,gebru2021datasheets}, and supply-chain
provenance~\citep{slsa} to agent skills. These lines pursue identity and behavior separately;
we add their composition.

%% file: sections/limitations.tex
\section{Scope and limitations}
\label{sec:limitations}

The limitations below define where the fingerprint applies.

\paragraph{Artifact similarity, and the adaptive attacker.}
The fingerprint compares what a skill \emph{is}, not what it \emph{does} at runtime, and
Section~\ref{sec:twoaxis} measures that boundary. No rewrite was optimized against it, though
the planes are public by design. Pushing one component's code below $0.85$ takes flipping about
$48$ of $320$ bits (the overall score can need more or fewer) while keeping the skill
functional, which is untested; keyed per-deployment planes would raise that cost but break
cross-organization comparability. Under the admission policy of Section~\ref{sec:twoaxis},
evasion sends a skill to the full audit and mimicry cannot shrink a check, since exact digests
set where it starts, leaving exposure in deduplication, families, and lineage records. Code
normalization also strips comments, so a comment-borne instruction leaves the code fingerprint
unchanged; digests still catch it.

\paragraph{Bounded by the encoder, the comparison set, and coverage.}
Independent cross-language equivalence is unlikely to be recovered, as one algorithm in two
languages embeds differently (not evaluated; a multilingual code
encoder~\citep{guo2022unixcoder} is the remedy, Appendix~\ref{app:crosslang}). Our method ladder also omits token-$n$-gram MinHash (which
recovers $19\%$ of semantic clones on SkillReuse's benchmark~\citep{zhu2026skillreuse}) and
code-specialized encoders. Tools appear in only $28\%$
of constructed skills, so the tools AUC of $1.000$ rests on a minority of pairs. The component vocabulary is fixed a priori, and we
compared the triple neither against a same-budget whole-skill code nor against a fused score,
so whether the split discriminates better, and whether three is the right number, remain
open.

\paragraph{Evaluation scale and labels.}
Our constructed results are on $100$ skills; the community check has $13$ positive pairs,
read through low-scoring pairs and a same-vendor judge (Appendices~\ref{app:community},
\ref{app:judge}); the adversarial benchmark has $50$ pairs and $10$ fingerprint negatives; and
the injection lineage results come almost entirely from generated injections.
Still owed are independent real-world lineage (fork histories, human-annotated families) and
an end-to-end measurement of the admission policy.

%% file: sections/conclusion.tex
\section{Conclusion}
\label{sec:conclusion}

Agent skills need an identity that survives paraphrase, renaming, and refactoring, which
cryptographic hashing is built to destroy. The per-component SimHash fingerprint, at most
$120$ bytes, tracks the embedding's ranking to within $0.03$ AUC at $77\times$ less storage on
a constructed corpus. The contribution is not the compression: per-component similarity
for skills has precedent~\citep{zhu2026skillreuse}; what we add is a compact code
per component, free of any corpus-fitted basis, that a verifier can index, compare at an a-priori threshold,
and read for which part of a skill moved. Identity is one axis of trust, not the whole of it.
The fingerprint recognizes a tampered copy of a registered skill but cannot decide whether it
is safe, so it composes with behavioral verification. A verdict is only as
durable as the identity it attaches to, which argues for family-level identity, in place of
byte equality, as the primitive a verifier builds on.

The \emph{convergent transformation} at its core plausibly fits models fingerprinted from
their outputs~\citep{yang2024llmfingerprint,nasery2025fingerprint}, untested here. Cryptographic hashing answers whether an artifact changed. A
locality-preserving per-component fingerprint answers whether it is the same artifact.

%% file: sections/appendix.tex
\appendix
\setcounter{table}{0}
\renewcommand{\thetable}{A\arabic{table}}
\renewcommand{\theHtable}{A\arabic{table}}
\setcounter{figure}{0}
\renewcommand{\thefigure}{A\arabic{figure}}
\renewcommand{\theHfigure}{A\arabic{figure}}

\section{A key to the reported AUCs}
\label{app:metricmap}
Several AUCs appear in this paper under different corpora, aggregations, and ground truths,
and they are not interchangeable. Table~\ref{tab:metricmap} is the key. The headline number
is the first row; every other row answers a narrower question. Two in particular are easy
to misread. The near-saturated adversarial AUCs reflect trivially separable cross-source
negatives, so detection at the operating threshold is the honest metric there
(Section~\ref{sec:robustness}). The low community-corpus overall AUC coincides with
name-based labels that inspection suggests often conflate distinct skills
(Appendix~\ref{app:community}).

\begin{table}[ht]
  \centering
  \small
  \caption{Every AUC reported in the paper, with the corpus and ground truth it was
  measured against. ``Micro-benchmark'' denotes the aggressive single-component rewrites of
  Table~\ref{tab:ladder}, reported per modality instead of as a weighted overall score.
  $^{\dagger}$The re-embedded fingerprint AUC is $0.931$ under the production engine's
  projection planes and $0.926$ under the analysis scripts' planes (Tables~\ref{tab:quantbaseline}
  and~\ref{tab:ablation}): two random draws of the same construction, inside one bootstrap
  interval. $^{\ddagger}$Original run, code preprocessing unrecorded; the re-embedding row uses
  AST-normalized code, and the gap between them is confined to the code component
  (Section~\ref{sec:fidelity}).}
  \label{tab:metricmap}
  \setlength{\tabcolsep}{4pt}
  \begin{tabular}{llll}
    \toprule
    AUC & Corpus (pairs) & Aggregation & Ground truth \\
    \midrule
    $\mathbf{0.974}^{\ddagger}$ & constructed, $100$ skills ($4{,}950$) & overall (weighted) & designed lineage \\
    $0.959/0.997/1.000$ & constructed ($4{,}950$) & prompt / code / tools & designed lineage \\
    $0.999$ & constructed ($4{,}950$) & TF-IDF baseline & designed lineage \\
    $0.992$--$0.996$ & micro-benchmark & per modality & single-component rewrite \\
    $0.993$ & micro-benchmark & embedding cosine & single-component rewrite \\
    $1.000 / 0.998$ & adversarial ($50$ pos.) & fingerprint / TF-IDF & transform pairs \\
    $0.739$ & OpenClaw, $1{,}000$ skills ($13$ pos.) & overall & name/hash duplicates \\
    $0.992/0.999/0.998$ & OpenClaw ($13$ pos.) & prompt / code / tools & name/hash duplicates \\
    $0.973$ & constructed, leave-family-out & overall & held-out family \\
    $0.931^{\dagger}$ vs.\ $0.959$ & constructed, re-embedded & fingerprint vs.\ cosine & designed lineage \\
    \bottomrule
  \end{tabular}
\end{table}

\section{The registry built on the fingerprint}
\label{app:registry}
Section~\ref{sec:decomposition} states that the same per-component distances scale to a
registry. Four pieces make that concrete.

\paragraph{Indexing.}
Each component is indexed independently, so a query skill can ask, per component, ``which
known skills are closest here?'' A component code is exactly the input a binary (Hamming)
index consumes: the store is a bit-packed array of $40$ bytes per skill per component,
$40$\,MB for a million skills, queried with the same exclusive-or and population-count
kernel as the pairwise comparison. An exact binary-flat scan answers a query in $0.15$\,ms
over $1{,}000$ skills and roughly $200$\,ms over $1$ million;\footnote{In our FAISS build the
binary-flat path returned distances inconsistent with an exact scan at code widths $\ge 256$
(we did not isolate the cause), so the prototype computes the same Hamming kernel in numpy.
This changes neither the asymptotics nor the results.} a binary HNSW or IVF
index~\citep{johnson2019faiss}, or the bit-block routing of~\citet{manku2007}, makes it
sub-linear on the $320$-bit codes unchanged. The composite index reduces the per-component
results three ways: a \emph{union} ranked by the best-matching component (recall), an
\emph{intersection} ranked by the worst (strict clone lookup), and per-component top-$k$.

\paragraph{Score normalization.}
Raw SimHash similarity is not interpretable as a percentage because of the floor of
Section~\ref{sec:theory}: unrelated skills sit near $0.65$, not $0.5$. The registry
normalizes each component's raw score $s$ against an empirical negative distribution with
mean $\mu$ and standard deviation $\sigma$, reporting $\Phi\big((s-\mu)/\sigma\big)$ with
$\Phi$ the standard normal CDF, so that $0.5$ denotes the random baseline, $0.84$ roughly one standard
deviation of signal, and $0.98$ roughly two. This is a per-component rescaling that makes
scores comparable across components and corpus snapshots; it does not convert a similarity
into a probability of any behavioral property. Statistics are versioned with each snapshot,
and because ours are fit on the evaluation corpus itself, thresholds on the normalized scale
should be revalidated on held-out skills before deployment.

\paragraph{Relationship labels.}
Table~\ref{tab:patterns} comes from a fixed rule list over the component scores $(p, c, t)$,
with high $= 0.80$, low $= 0.60$, and a component missing from either skill scoring $0$,
applied in priority order: \emph{clone} if $p, c \ge 0.80$; \emph{reimplemented} if
$p \ge 0.80$ and $c < 0.60$; \emph{repurposed} if $p < 0.60$ and $c \ge 0.80$; \emph{tool swap}
if $p \ge 0.80$ or $c \ge 0.80$, and $t < 0.60$; \emph{variant} if $p, c \ge 0.60$;
\emph{unrelated} if $p, c < 0.60$; otherwise the fallback \emph{mixed}, outside the six named
labels. Because clone is checked first, a tools-only change to an otherwise identical skill
$(1, 1, 0)$ is labeled clone. Coverage and asymmetry are reported with each comparison but are
not inputs. Because a missing tools component scores $0$, the
tool-swap rule also fires for pairs in which neither skill declares tools, so that label
overstates genuine tool swaps.

\paragraph{SkillBOM.}
Because the fingerprint is constant-size and component-structured, it serializes naturally
into a bill of materials. We emit a CycloneDX-shaped \emph{SkillBOM}~\citep{cyclonedx} in
which each skill records its per-component fingerprint (tagged by component), its family and
canonical sibling, its novelty score, a provenance block, and a reserved slot for behavioral
attestations. The format is designed to carry a cryptographic digest per component beside
each fingerprint, signable with standard supply-chain tooling~\citep{newman2022sigstore}, so
that one document records both exact integrity (the digests) and lineage (the fuzzy
fingerprint). Our prototype emits the fingerprints but not yet the digests, which the
admission policy of Section~\ref{sec:twoaxis} requires. The two are complements, not
substitutes.
It follows the documentation-as-artifact lineage of model cards and
datasheets~\citep{mitchell2019modelcards,gebru2021datasheets} and the supply-chain
provenance frameworks for conventional software~\citep{slsa}, extended to the agent-skill
artifact. Being hash-only, it can be shared without shipping skill source, which is what
makes a cross-organization registry tractable. Sharing is not confidentiality, though: the
planes are public and deterministic, so anyone holding a candidate skill can fingerprint it
and test for a match.

\paragraph{An optional triage gateway.}
When a policy decision is wanted rather than a lookup, the same components feed a gateway
that fingerprints an incoming skill, searches the registry, and resolves against curated
families: a strong normalized match to a trusted family routes to a behavioral check that
\emph{starts} from the components whose digests differ from the family's verified member and
covers their interactions, a strong
match to a flagged family to \emph{priority review}, and mixed or novel signals to a
\emph{full audit} with the per-component evidence attached. The gateway is a router, not an
oracle: no fuzzy match allows or blocks a skill on its own, and a skill whose code matches a
flagged family but whose prompt looks benign is sent to review \emph{with evidence} instead
of being silently allowed or silently blocked.

\paragraph{Family recovery and novelty, in full.}
Section~\ref{sec:decomposition} reports the headline recovery numbers. Two further
measurements support them. Of each skill's four genuine variants, $88.8\%$ appeared in its
top four neighbors, so the structure is recovered at rank rather than only at the single
nearest neighbor. For the withheld families, members' best cross-family match fell to $0.760$
against a same-family support of $0.932$, giving a novel-versus-known separability AUC of
$0.973$ and leaving the $0.85$ operating threshold comfortably inside the margin.

\paragraph{The false-clone pairs on real skills.}
The $11$ pairs flagged as clones among the $125{,}678$ distinct-name pairs of
Section~\ref{sec:decomposition} are reuse families, not errors. Two representative cases are
a builder and its base (\texttt{mcp-builder} and \texttt{mcp}, $0.92$) and a workflow
variant and its underlying tool (\texttt{git-advanced-workflows} and \texttt{git}, $0.91$).
In both, the triple attributes the match to the component the two skills genuinely share,
which is what lets a registry present the pair for review with evidence instead of asserting
a duplicate.

\section{Bit budget and concentration}
\label{app:bitbudget}
The bit-budget claims of Section~\ref{sec:theory} follow from the SimHash estimator's
variance $p(1-p)/B$ with $p = \arccos(c)/\pi$. Beyond the $1/\sqrt{B}$ standard deviation,
the estimator \emph{concentrates} rather than merely averaging: as a mean of $B$
independent values in $[0,1]$, a Hoeffding bound gives
\begin{equation}
  \Pr\!\big[\, \lvert \mathrm{sim} - \mathbb{E}[\mathrm{sim}] \rvert > t \,\big]
  \;\le\; 2\,e^{-2 B t^{2}},
  \label{eq:hoeffding}
\end{equation}
so at $B = 320$ a deviation larger than $0.1$ has probability below
$2e^{-6.4} \approx 0.003$. The fidelity is therefore a high-probability guarantee on each
comparison, not just a low-variance tendency in aggregate. Table~\ref{tab:bits} traces the
consequence across bit budgets: five banks of $64$ bits is the knee of the curve.

\begin{table}[ht]
  \centering
  \small
  \caption{SimHash bit budget. $\sigma$ is the projection noise $\sqrt{p(1-p)/B}$ of one pair
  at the mean positive cosine, falling as $1/\sqrt{B}$; Gap$/\sigma$ divides the gap predicted
  from the mean cosines ($0.101$) by the combined projection noise of a positive and a negative
  pair. Both exclude the spread across pairs, whose overlap keeps the cosine AUC at $0.993$. AUC converges to the cosine ceiling
  ($0.993$); the $B{=}640$ value exceeds it by sampling noise. Five $64$-bit banks
  ($B = 320$) is the operating point.}
  \label{tab:bits}
  \begin{tabular}{lccc}
    \toprule
    Configuration & $\sigma$ & Gap$/\sigma$ & AUC (obs.) \\
    \midrule
    $1\times64$ ($B{=}64$)   & $0.052$ & $1.3$ & $0.917$ \\
    $\mathbf{5\times64}$ ($B{=}320$) & $\mathbf{0.023}$ & $\mathbf{2.9}$ & $\mathbf{0.992}$ \\
    $10\times64$ ($B{=}640$) & $0.016$ & $4.1$ & $0.998$ \\
    $\infty$ (cosine)        & $\to 0$ & $\to\infty$ & $0.993$ \\
    \bottomrule
  \end{tabular}
\end{table}

\section{Quantization fidelity and alternative compressions}
\label{app:quant}

\begin{table}[ht]
  \centering
  \small
  \caption{Method ladder on aggressive single-component rewrites
  (Section~\ref{sec:fidelity}). The embedding supplies rewrite-invariance; multi-bank SimHash
  preserves it at $77\times$ compression. AUC ranges span the prompt and code modalities
  where measured separately.}
  \label{tab:ladder}
  \begin{tabular}{lcl}
    \toprule
    Method & AUC & Note \\
    \midrule
    Raw CTPH (ssdeep-style)            & $0.53$--$0.61$ & no spatial locality in skills \\
    AST-normalized CTPH                & $0.63$ & Type-2 clones only \\
    Constrained paraphrase $+$ SimHash & $0.89$ & LLM noise $>$ convergence benefit \\
    Embedding cosine                   & $0.99{+}$ & accurate but kilobyte-scale \\
    \textbf{Multi-bank SimHash ($5\times64$)} & $\mathbf{0.992}$--$\mathbf{0.996}$
      & matches cosine at $77\times$ compression \\
    \bottomrule
  \end{tabular}
\end{table}

The comparison the fingerprint is built for is against the full-precision embedding cosine
it approximates. Re-embedding the corpus and clustering on the raw $768$-dimensional cosine
versus the $320$-bit fingerprint computed from the \emph{same} vectors, the fingerprint
reproduces cosine to within $0.004$ AUC on every component (prompt $0.971$ versus $0.975$,
code $0.991$ versus $0.994$, tools $1.000$ for both) and to within $0.03$ overall ($0.931$
versus $0.959$), at $77\times$ less storage. Skill-level bootstrap CIs overlap throughout
(for example prompt $[0.937, 0.998]$ versus $[0.934, 0.998]$). The $320$-bit signature
loses bits, not the metric. This comparison uses the production engine's projection planes;
Tables~\ref{tab:quantbaseline} and~\ref{tab:ablation} use the analysis scripts' planes, a
second random draw that gives $0.926$ overall, inside the same bootstrap interval
($[0.899, 0.982]$).

\label{app:compression}
The fingerprint is not, however, the most accurate way to compress the embedding, and it
should not claim to be. Table~\ref{tab:quantbaseline} compares it on the same corpus
against float16, int8, a $64$-dimensional PCA, and an $8\times16$ product
quantizer~\citep{jegou2011pq}. On accuracy alone a corpus-fit PCA matches or beats every
other representation (its in-sample AUC is an optimistic upper bound), and even a $4$-byte
product quantizer edges the fingerprint. The fingerprint's advantage is operational: it is
the only representation here that is simultaneously a constant-size, self-contained bit
string searched by exclusive-or and population-count, with no shared basis or codebook that
every consumer must distribute and keep synchronized, and with per-component bit
attribution. Portability has a cost of its own, which belongs in the same comparison: codes
are comparable only between parties on the same encoder version, so the fingerprint trades a
shared basis or codebook for a shared encoder (Appendix~\ref{app:ablation}). The difference is
that an encoder version is a public, versioned artifact identical for every consumer, whereas
a PCA basis or product-quantization codebook is fit to one party's corpus and must be
redistributed whenever it is refit. For a portable, append-only, cross-organization registry
that is the right trade, at a cost of about $0.03$--$0.04$ AUC against full cosine.

\begin{table}[ht]
  \centering
  \small
  \caption{The fingerprint against other compressions of the same $768$-d embedding
  ($100$-skill corpus; overall AUC and 1-NN purity against the $20$-group ground truth).
  SimHash is competitive at the smallest size without a corpus-fitted basis and is the only Hamming-indexable,
  codebook-free option. $^{*}$PCA is fit in-sample (optimistic) and needs a shared basis.}
  \label{tab:quantbaseline}
  \begin{tabular}{lccccc}
    \toprule
    Representation & bytes/comp & 1-NN & AUC & Hamming idx. & codebook-free \\
    \midrule
    cosine f32        & $3072$ & $0.94$ & $0.959$ & no & yes \\
    float16           & $1536$ & $0.94$ & $0.959$ & no & yes \\
    int8              & $768$  & $0.94$ & $0.959$ & no & yes \\
    PCA-64            & $256$  & $1.00$ & $1.000^{*}$ & no & no \\
    product quant.\ ($8\times16$) & $4$ & $0.87$ & $0.948$ & no & no \\
    \textbf{SimHash $320$\,b (ours)} & $\mathbf{40}$ & $0.93$ & $0.926$ & \textbf{yes} & \textbf{yes} \\
    \bottomrule
  \end{tabular}
\end{table}

\section{Encoder ablation and ingestion cost}
\label{app:ablation}
Because the encoder is the most important choice in the system, we vary it across four
hosted encoders of different generations and dimensions, with identical preprocessing
(Table~\ref{tab:ablation}); open encoders such as E5 and CodeBERT were unavailable in our
environment. Projection planes come from NumPy's \texttt{default\_rng} with seeds $1$, $2$,
$3$ for prompt, code, and tools. The encoder matters before compression: cosine AUC ranges
from $0.934$ for the multilingual encoder (weakest on code) to $0.976$ for the newest and
$4\times$ widest. After compression the fingerprint spans only $0.919$--$0.935$, and the
ordering is not preserved (the multilingual encoder is weakest in cosine but not as a
fingerprint), so at $320$ bits the hash absorbs most of a stronger encoder's gain. That
accuracy is not free at ingestion, because the
widest encoder costs $4\times$ the embedding bytes and ${\approx}\,30\%$ more latency per
call. The cost is paid once: the fingerprint is $40$ bytes per component and the lookup is
$O(1)$ Hamming \emph{regardless of the encoder}, so a registry's storage and query cost do
not change with the encoder, only its embedding bill does. The fingerprint tracks each
encoder's cosine to within $0.01$--$0.04$, so a stronger encoder can be swapped in without
changing the hash, the signature size, or the index, though it buys less after compression
than its cosine suggests. The ablation stays
within one vendor family; an open-encoder comparison is left to an environment that can
host them.

\begin{table}[ht]
  \centering
  \small
  \caption{Encoder ablation on the $100$-skill corpus: quality and ingestion cost. Latency
  is the median of $15$ single-text calls including network. Per-component cosine AUC follows the
  overall cosine ordering (tools is $1.000$ throughout).}
  \label{tab:ablation}
  \begin{tabular}{lccccc}
    \toprule
    Encoder & dim & embed (kB) & latency (ms) & fp (B) & AUC (cos / fp) \\
    \midrule
    text-embedding-004 & $768$ & $3.0$ & $170$ & $40$ & $0.953$ / $0.919$ \\
    \textbf{text-embedding-005} (default) & $768$ & $3.0$ & $173$ & $40$ & $0.959$ / $0.926$ \\
    text-multilingual-embedding-002 & $768$ & $3.0$ & $151$ & $40$ & $0.934$ / $0.927$ \\
    gemini-embedding-001 & $3072$ & $12.3$ & $228$ & $40$ & $\mathbf{0.976}$ / $\mathbf{0.935}$ \\
    \bottomrule
  \end{tabular}
\end{table}

\section{Cross-language, by scenario}
\label{app:crosslang}
We claim robustness to cross-language rewriting and also list cross-language equivalence as
a limitation; these are not in conflict, and Table~\ref{tab:crosslang} makes the boundary
exact. Under \emph{controlled translation} the prompt is unchanged, so overall identity is
recovered (${\approx}\,0.90$, above the $0.85$ threshold) even though the translated code
component degrades to ${\approx}\,0.82$: the prompt carries it while the code signal
weakens. An independently authored reimplementation in another language, which shares
neither the prompt nor the code surface, was not evaluated because we lack a labeled
independent-multilingual corpus; we expect it to fail and list it as a limitation.

\begin{table}[ht]
  \centering
  \small
  \caption{Cross-language, by scenario. Controlled translation is measured; independent
  multilingual reimplementation shares no component and was not evaluated.}
  \label{tab:crosslang}
  \begin{tabular}{lcccl}
    \toprule
    Scenario & prompt & code & overall & outcome \\
    \midrule
    Translate to JavaScript & $1.000$ & $0.823$ & $0.904$ & recovered (via prompt) \\
    Translate to TypeScript & $1.000$ & $0.819$ & $0.902$ & recovered (via prompt) \\
    Translate to Go         & $1.000$ & $0.818$ & $0.901$ & recovered (via prompt) \\
    \midrule
    Independent multilingual reimpl. & n/a & n/a & n/a & not evaluated (expected failure) \\
    \bottomrule
  \end{tabular}
\end{table}

\section{Out-of-distribution check on a community corpus}
\label{app:community}
The harder test is real skills with noisy labels. We applied the fingerprint to
the OpenClaw community skills repository~\citep{openclaw}, $45{,}266$ community-contributed skills,
sampling $1{,}000$ ($611$ with code, $389$ prompt-only). Ground truth here is weak by
construction: $12$ name-based duplicate groups plus $3$ content-hash groups give only $13$
positive pairs against $5{,}000$ sampled negatives, and ``same name'' does not imply ``same
skill'' when skills are authored independently. Per-component fingerprints stayed consistent
across this check: prompt AUC $0.992$, code $0.999$, tools $0.998$, nearly unchanged from their
constructed-corpus values. With only $13$ positive pairs this is a consistency check, not
a generalization claim.

The overall AUC, however, fell to $0.739$, and inspection points to the labels. The positive
pairs are bimodal: true clones cluster at $0.85$--$1.00$ (two
exact, at $1.000$), while six ``duplicates'' that merely share a name fall at
$0.27$--$0.53$. If those six are distinct skills, as the clearest case suggests, a label set
calling all $13$ pairs positive is roughly half wrong.
Two independently written \texttt{prompt-optimizer} skills score $0.265$: one does iterative
refinement, the other single-pass style transfer. Together these suggest that the
fingerprint tracks implementation more closely than names do. Because the reading comes from
inspecting pairs the fingerprint scored low, human annotation of the $13$ pairs would be
needed to confirm it; the LLM-judge spot check below points the same way but shares the
encoder's vendor family. Section~\ref{sec:limitations} records the limit: $13$ positive
pairs cannot support a generalization claim.

\section{Agreement with an LLM judge}
\label{app:judge}
\begin{figure}[ht]
  \centering
  \includegraphics[width=0.62\linewidth]{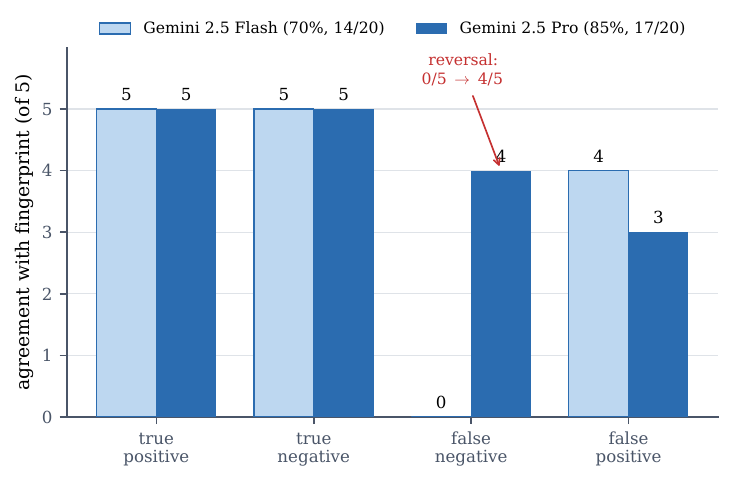}
  \caption{Agreement with the fingerprint by category, per judge. Both judges agree on the
  easy cells. The stronger judge gains in the false-negative cell (same-name,
  different-implementation pairs, $0/5$ to $4/5$) and loses one in the false-positive cell
  ($4/5$ to $3/5$).}
  \label{fig:judge}
\end{figure}

As a sanity check, not core evidence, we compare the fingerprint against a competent
reader. We sampled $20$ pairs from the SkillsBench benchmark~\citep{li2026skillsbench}
across four cells (true positive, true negative, and the two hard cells:
same-name/different-implementation and different-name/similar-implementation) and elicited
an independent similarity judgment from two judges of different strength, following
standard LLM-as-judge practice~\citep{zheng2023judge} and noting its
biases~\citep{wang2023unfair}. Agreement with the fingerprint rose with judge strength,
from $70\%$ ($14/20$) for the weaker judge to $85\%$ ($17/20$) for the stronger
(Figure~\ref{fig:judge}). The net gain of three is $+4$ in the
same-name/different-implementation cell, where the stronger judge flipped from $0/5$ to
$4/5$ agreement, and $-1$ in the different-name/similar-implementation cell ($4/5$ to $3/5$). We read the direction as suggestive: one would not expect a stronger
reasoner to agree \emph{more} if the signal were purely lexical, which is consistent with
the code component capturing implementation-level differences. At $20$ pairs with
overlapping Wilson intervals ($[0.48, 0.85]$ for $70\%$ and $[0.64, 0.95]$ for $85\%$), and with both judges
sharing a vendor family with the embedding, this is a sanity check rather than a powered or
independent result, and we do not count it among the paper's findings.

\section{Supplementary figures}
\label{app:figures}
\begin{figure}[ht]
  \centering
  \includegraphics[width=0.80\linewidth]{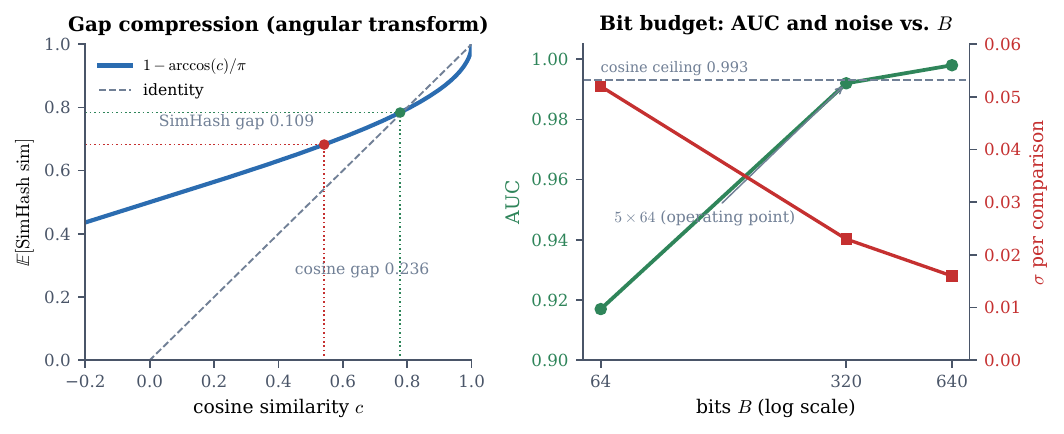}
  \caption{Why the bits preserve identity. \emph{Left:} the transfer function
  $1-\arccos(c)/\pi$ has slope $0.38$--$0.51$ over the observed cosine range, compressing the
  positive/negative gap from $0.236$ (cosine)
  to $0.109$ ($320$ bits); the negative floor near $0.65$ comes from the positive background
  cosine of unrelated components (mean $0.54$), not from the hash. \emph{Right:}
  per-comparison noise $\sigma$ falls as $1/\sqrt{B}$ while AUC approaches the cosine ceiling
  ($0.993$); the $5\times64$ operating point sits at the knee.}
  \label{fig:theory}
\end{figure}

\begin{figure}[ht]
  \centering
  \includegraphics[width=0.66\linewidth]{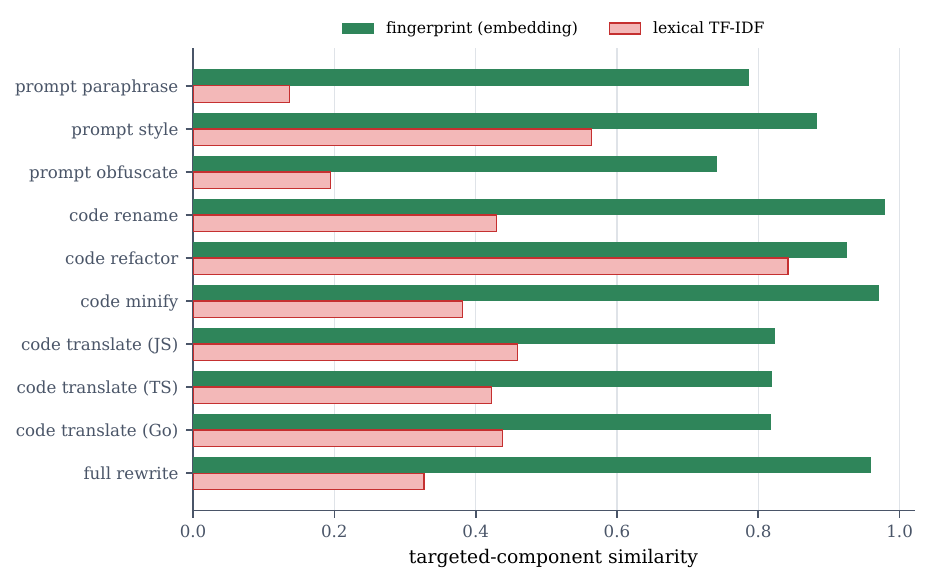}
  \caption{Targeted-component similarity across ten rewrite transforms. The fingerprint's
  targeted component drops to $0.74$--$0.98$ while the untouched component stays at $1.0$,
  which keeps overall similarity above the $0.85$ threshold on every transform
  (Table~\ref{tab:adv}). The full rewrite edits both components; its bar shows code
  ($0.96$), and its prompt scores $0.79$.
  A lexical TF-IDF baseline collapses on the transforms that destroy surface text.}
  \label{fig:adversarial}
\end{figure}

\begin{figure}[ht]
  \centering
  \includegraphics[width=0.78\linewidth]{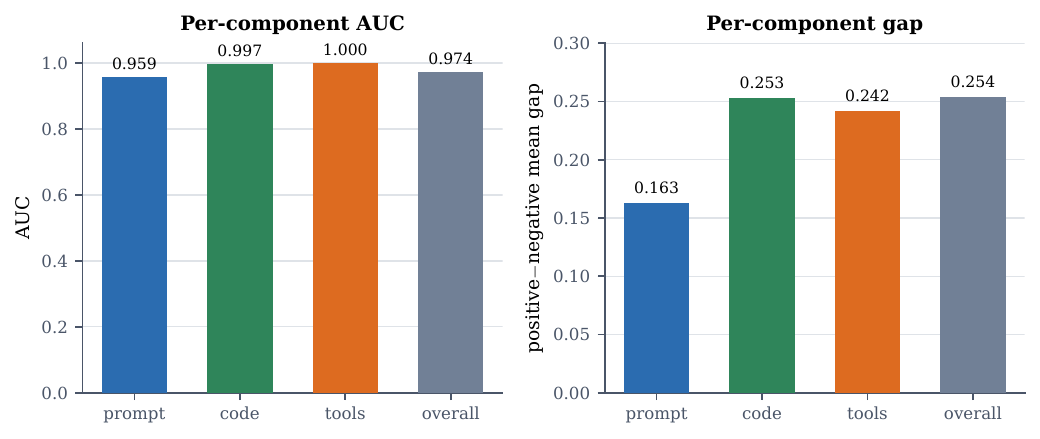}
  \caption{Per-component discrimination. Code and tools separate skills more sharply than
  prompt, by both AUC (left) and positive/negative mean gap (right); the overall column
  aggregates component similarities (Equation~\ref{eq:overall}) and hides this spread.}
  \label{fig:percomp}
\end{figure}

\begin{figure}[ht]
  \centering
  \begin{minipage}[t]{0.48\linewidth}
    \centering
    \includegraphics[width=\linewidth]{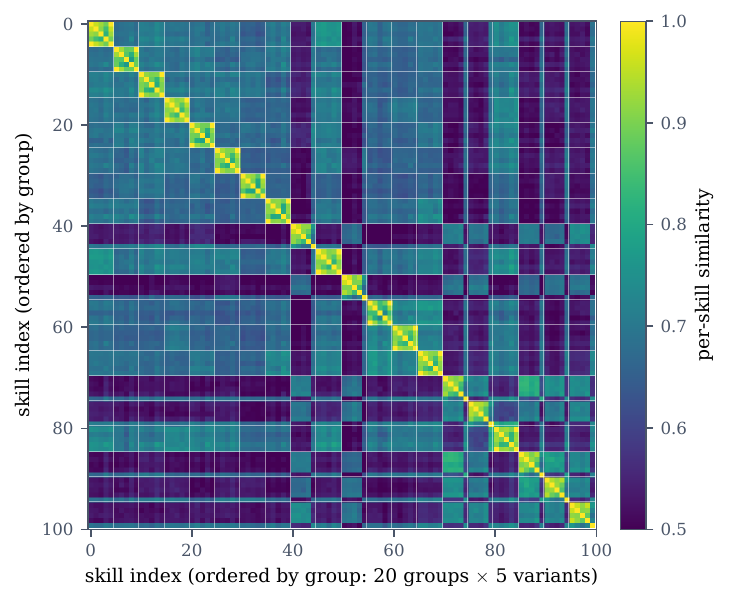}
    \caption{Pairwise similarity over the $100$-skill corpus, ordered by group.}
    \label{fig:heatmap}
  \end{minipage}\hfill
  \begin{minipage}[t]{0.48\linewidth}
    \centering
    \includegraphics[width=\linewidth]{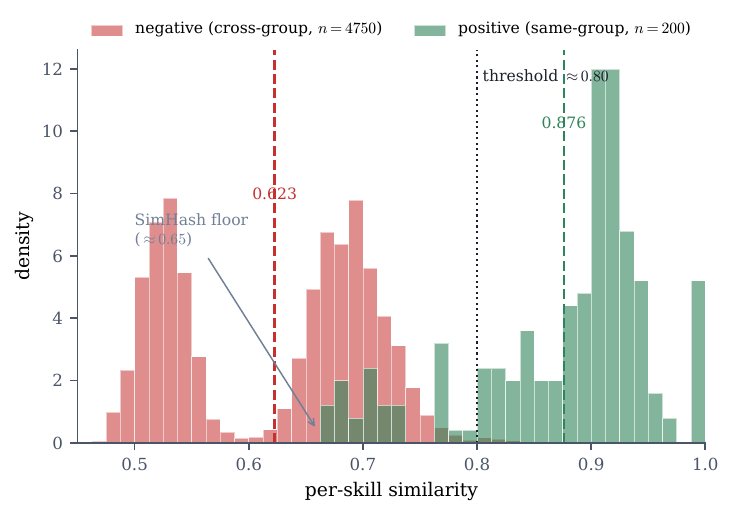}
    \caption{Positive and negative score distributions on the same corpus. The dashed line
    is the $0.80$ clustering threshold of Section~\ref{sec:decomposition}; the operating
    threshold is $0.85$.}
    \label{fig:scoredist}
  \end{minipage}
\end{figure}

\begin{figure}[ht]
  \centering
  \includegraphics[width=0.66\linewidth]{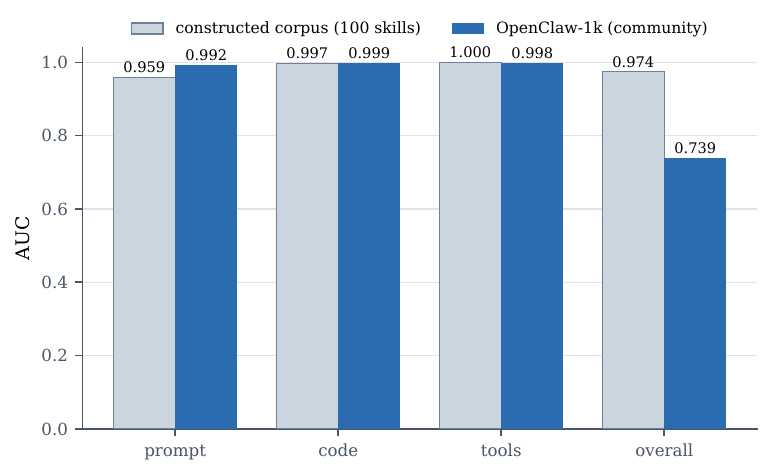}
  \caption{Per-component AUC on the $1{,}000$-skill community corpus against the
  constructed corpus. Per-component discrimination is stable; the drop is confined to the
  overall score, which is scored against name-based labels, several of which conflate
  distinct skills.}
  \label{fig:ood}
\end{figure}

Figure~\ref{fig:theory} plots the two effects analyzed in Section~\ref{sec:theory}.
Figure~\ref{fig:adversarial} plots the per-transform detail behind
Table~\ref{tab:adv}. Figure~\ref{fig:percomp} visualizes the per-component spread of
Table~\ref{tab:percomp}. Figure~\ref{fig:heatmap} shows the $100\times100$ similarity matrix
of the fidelity corpus of Section~\ref{sec:fidelity}: $5\times5$ diagonal blocks, plus
off-diagonal stripes produced by Equation~\ref{eq:overall}'s missing-component penalty
(cross-group pairs average $0.53$ when only one skill declares tools, against $0.69$--$0.72$
when both or neither do). Figure~\ref{fig:scoredist} shows the corresponding positive
and negative score distributions, whose separation and compressed gap are what
Section~\ref{sec:theory} predicts. Figure~\ref{fig:ood} shows the per-component AUCs on the
community corpus, nearly unchanged from their constructed-corpus values even as the
overall AUC falls to $0.739$ under noisy name-based labels.

